# Broadband calibration-free cavity-enhanced complex refractive index spectroscopy using a frequency comb

ALEXANDRA C. JOHANSSON,[1] LUCILE RUTKOWSKI,[1] ANNA FILIPSSON,[1] THOMAS HAUSMANINGER,[1] GANG ZHAO,[1,2] OVE AXNER,[1] AND ALEKSANDRA FOLTYNOWICZ[1,*]

[1]*Department of Physics, Umeå University, 901 87 Umeå, Sweden*
[2]*State Key Laboratory of Quantum Optics and Optics Devices, Institute of Laser Spectroscopy, Shanxi University, Taiyuan 030006, China*
*\*aleksandra.foltynowicz@umu.se*

**Abstract:** We present broadband cavity-enhanced complex refractive index spectroscopy (CE-CRIS), a technique for calibration-free determination of the complex refractive index of entire molecular bands via direct measurement of transmission modes of a Fabry-Perot cavity filled with the sample. The measurement of the cavity transmission spectrum is done using an optical frequency comb and a mechanical Fourier transform spectrometer with sub-nominal resolution. Molecular absorption and dispersion spectra (corresponding to the imaginary and real parts of the refractive index) are obtained from the cavity mode broadening and shift retrieved from fits of Lorentzian profiles to the individual cavity modes. This method is calibration-free because the mode broadening and shift are independent of the cavity parameters such as the length and mirror reflectivity. In this first demonstration of broadband CE-CRIS we measure simultaneously the absorption and dispersion spectra of three combination bands of $CO_2$ in the range between 1525 nm and 1620 nm and achieve good agreement with theoretical models. This opens up for precision spectroscopy of the complex refractive index of several molecular bands simultaneously.

2018-05-02

## 1. Introduction

Transmission modes of a Fabry-Perot cavity provide high sensitivity to intracavity absorption and dispersion because their center frequency, width, and amplitude are modified near molecular transitions. Most cavity-enhanced spectroscopic techniques provide information about molecular absorption either by measuring the intensity change in cavity transmission, as in e.g. cavity-enhanced absorption spectroscopy (CEAS) [1], optical feedback CEAS [2], and off-axis integrated cavity output spectroscopy [3], or by measuring the decay time of the light leaking out of the cavity, as is done in cavity ring-down spectroscopy (CRDS) [4, 5]. Of these, only CRDS is calibration-free, while the other techniques require an independent measurement of the cavity finesse to quantify the absorption. Simultaneous information about molecular absorption and dispersion is provided by noise-immune cavity-enhanced optical heterodyne molecular spectroscopy (NICE-OHMS), which combines cavity enhancement with frequency modulation (FM) [6, 7]. However, because of the use of the FM scheme, NICE-OHMS requires calibration using samples with a known concentration.

In contrast to the amplitude change, the shift of the center frequency and the broadening of the cavity modes are directly proportional to the real and the imaginary parts of the molecular index of refraction, respectively, but independent of the cavity parameters, e.g. mirror reflectivity and cavity length. The cavity mode broadening caused by molecular absorption was first observed by Nakagawa *et al.* [8]. Later on, Long *et al.* [9] retrieved a high-precision absorption spectrum of a $CO_2$ transition from the cavity mode broadening measured using frequency agile rapid scanning spectroscopy (FARS) [10]. Hodges *et al.* [11] demonstrated that complementary information about molecular dispersion can be extracted from the cavity mode shift. Cygan *et al.* developed these concepts into two techniques: cavity mode width spectroscopy (CMWS) [12] and one-dimensional cavity mode-dispersion spectroscopy (1D-CMDS) [13]. Measuring the broadening and shift of the cavity resonances simultaneously allows simultaneous and calibration-free assessment of the imaginary and real parts of the complex refractive index of molecular transitions [14, 15]. However, cavity-enhanced complex refractive index spectroscopy (CE-CRIS) [16] performed using continuous-wave lasers has a limited spectral coverage and requires precise sequential scanning of a narrow linewidth laser over individual cavity modes, which is time consuming even for FARS. Therefore, existing demonstrations of CE-CRIS are limited to 1-2 individual molecular transitions. Moreover, in 1D-CMDS only the relative (rather than the absolute) cavity mode positions are measured to avoid the influence of the cavity length drift, while the absolute frequency calibration is obtained by referencing the probe laser to a frequency comb [14, 15].

Here we use a frequency comb to perform CE-CRIS over 15 THz of bandwidth by measuring the transmission spectrum of a cavity filled with $CO_2$ using a Fourier transform spectrometer (FTS) with sub-nominal resolution [17, 18]. The narrow lines of a stabilized frequency comb are simultaneously scanned across all cavity modes within the laser bandwidth and the intensities of the individual comb lines are measured by matching the nominal resolution of the spectrometer to the comb repetition rate [17, 18]. A Lorentzian function is fitted to each cavity mode to retrieve its center frequency, width, and amplitude. We have previously demonstrated that cavity mode frequencies can be used for high-precision retrieval of broadband intracavity group delay dispersion [19]. Here we utilize the full potential of cavity mode characterization and retrieve dispersion and absorption spectra of three combination bands of $CO_2$ from the cavity mode shifts and broadenings, as well as the corresponding cavity-enhanced transmission spectra from the cavity mode amplitudes. The entire measurement takes less than half an hour and the obtained spectra show good agreement with the corresponding models. This technique opens up for precision spectroscopy of the real and the imaginary parts of the complex refractive index, and calibration-free retrieval of line parameter of several molecular bands simultaneously.

## 2. Theory

The complex refractive index of a gas near a molecular transition can be expressed as [13]

$$n_a(\nu) = n + n'(\nu) - i\kappa(\nu), \tag{1}$$

where $n$ is the real non-resonant component, while $n'(\nu)$ and $\kappa(\nu)$ are the real and imaginary resonant components, respectively, directly proportional to the dispersion and absorption coefficients of the molecular transition, $\varphi(\nu)$ and $\alpha(\nu)$ [cm$^{-1}$],

$$n'(\nu) = \frac{1}{2k_0}\varphi(\nu), \tag{2}$$

and

$$\kappa(\nu) = \frac{1}{2k_0}\alpha(\nu), \tag{3}$$

respectively, where $k_0 = 2\pi\nu_0/c$ is the wave vector [cm$^{-1}$] corresponding to the transition frequency $\nu_0$ [Hz] and $c$ is the speed of light [cm/s]. The dispersion and absorption coefficients of the transition are related to the imaginary and the real part of the complex lineshape function $\chi$ (cm) as

$$\varphi(\nu) = -S'c_{rel}p\,\text{Im}\,\chi(\nu), \tag{4}$$

and

$$\alpha(\nu) = S'c_{rel}p\,\text{Re}\,\chi(\nu), \tag{5}$$

respectively, where $S'$ is the integrated molecular transition linestrength [cm$^{-2}$/atm], $c_{rel}$ is the relative concentration of the absorbing species (dimensionless), and $p$ is the sample pressure [atm].

The frequency dependent electric field transmission function of a Fabry-Perot cavity containing an absorbing medium can be expressed as [20]

$$h(\nu) = \frac{T\exp(-\alpha L/2)\exp(-i2\pi n\nu L/c - i\phi/2 - i\varphi L/2)}{1 - R\exp(-\alpha L)\exp(-i4\pi n\nu L/c - i\phi - i\varphi L)}, \tag{6}$$

where $T$ and $R$ are the frequency dependent intensity transmission and reflection coefficients of the cavity mirrors, respectively, $L$ is the cavity length [cm], and $\phi$ is the intracavity dispersion caused by the cavity mirror coatings. Note that while $\varphi$ and $\alpha$ have strong frequency dependence around a molecular transition, $T$, $R$, $n$ and $\phi$ are slowly varying functions of frequency. The intensity transmission function of the cavity, defined as $H(\nu) = h(\nu)h^*(\nu)$, is given by

$$H(\nu) = \frac{T^2 e^{-\alpha L}}{(1 - Re^{-\alpha L})^2}\frac{1}{1 + 4\dfrac{Re^{-\alpha L}}{(1 - Re^{-\alpha L})^2}\sin^2\left[2\pi n\nu\dfrac{L}{c} + \dfrac{\phi}{2} + \varphi\dfrac{L}{2}\right]}. \tag{7}$$

The cavity mode center frequencies, $\nu_k$, are obtained by equalizing the argument of the sine term in Eq. (7) with an integer multiple of $\pi$, which yields

$$\nu_k = \frac{c}{2Ln}\left(k - \frac{\phi}{2\pi}\right) - \frac{c}{4\pi n}\varphi = \nu_k^0 - \nu_\varphi, \tag{8}$$

where $\nu_k^0$ is the center frequency of the $k$th mode of the cavity in the absence of molecular transition and $\nu_\varphi$ is the cavity mode frequency shift caused by the dispersion of the molecular resonance,

$$\nu_\varphi(\nu) = \frac{c}{4\pi n}\varphi(\nu). \tag{9}$$

Close to a resonance of a high finesse cavity the sine term in Eq. (7) can be linearized around $(\nu - \nu_k)$, which reveals the Lorentzian form of the cavity modes, given by

$$H(\nu-\nu_k) \approx \frac{T_k^2 e^{-\alpha L}}{\left(1-R_k e^{-\alpha L}\right)^2} \frac{1}{1+\left[\dfrac{\nu-\nu_k}{\dfrac{c}{4\pi nL}\dfrac{1-R_k e^{-\alpha L}}{\sqrt{R_k e^{-\alpha L}}}}\right]^2}, \qquad (10)$$

where the mirror transmission and reflection coefficients, $T$ and $R$, have been given an index $k$ to account for their frequency dependence. The full width at half maximum of the Lorentzian cavity modes, $\Gamma_k$, is given by twice the term in the denominator of the expression in square brackets in Eq. (10), i.e.

$$\Gamma_k = 2\frac{c}{4\pi nL}\frac{1-R_k e^{-\alpha L}}{\sqrt{R_k e^{-\alpha L}}}. \qquad (11)$$

For a high-finesse cavity ($\sqrt{R_k} \approx 1$) and low single pass absorption ($\alpha L \ll 1$), this expression can be series expanded to

$$\Gamma_k = \frac{c}{2\pi nL}(1-R_k) + \frac{c}{2\pi n}\alpha = \Gamma_k^0 + \Gamma_\alpha, \qquad (12)$$

where $\Gamma_k^0$ is the cavity mode width in the absence of molecular transition and $\Gamma_\alpha$ is the mode broadening caused by the absorption of the molecular resonance,

$$\Gamma_\alpha(\nu) = \frac{c}{2\pi n}\alpha(\nu). \qquad (13)$$

The accuracy of Eq. (12) has been investigated in ref. [12], where it was shown that for α up to $16 \times 10^{-6}$ cm$^{-1}$ the relative error induced by the linearization is below $5 \times 10^{-4}$.

Finally, the cavity mode on-resonance transmission, $A_k^\alpha$, defined as $H(\nu-\nu_k=0)$, is given by

$$A_k^\alpha = \frac{T_k^2 e^{-\alpha L}}{\left(1-R_k e^{-\alpha L}\right)^2}. \qquad (14)$$

Note that while the mode shift and broadening caused by resonant molecular dispersion and absorption, i.e. $\nu_\varphi$ and $\Gamma_\alpha$, are independent of the cavity length and the cavity mirror transmission and reflectivity, and thus the mode number $k$, the on-resonance cavity transmission, $A_k^\alpha$, is not. Moreover, direct measurement of cavity transmission yields the power transmitted through the mode, $P_k = A_k^\alpha P_k^0$, where $P_k^0$ is the power of the comb line that probes the cavity mode, which needs to be known in order to retrieve $A_k^\alpha$.

## 3. Experimental Setup

The experimental setup, shown in Fig. 1, follows in large the one used in ref. [19]. The resonances of a Fabry-Perot cavity are probed by an Er:fiber frequency comb, and the light transmitted through the cavity is analyzed with a fast-scanning Fourier transform spectrometer (FTS) with sub-nominal resolution [18]. The cavity, which has a finesse of ~1700, consists of two dielectric mirrors glued to a 45-cm-long stainless steel tube, yielding a free spectral range (FSR) of 333 MHz. The cavity length is stabilized by the Pound-Drever-Hall (PDH) technique to a narrow-linewidth continuous-wave (cw) Er:fiber laser, which, in turn, is locked to the P8e $CO_2$ transition at 1576.9396 nm using sub-Doppler NICE-OHMS [21] based on the instrumentation described in ref. [22]. The cw laser is phase modulated using a fiber-coupled electro-optic modulator (EOM) and coupled into free space via a fiber-based optical circulator (OC), which is also used to pick up the cavity reflection for the PDH error signal detection. This error signal is fed into a proportional-integral controller, and the correction signal is sent to a piezo-electric transducer (PZT) placed between one of the cavity mirrors and the tube. The cavity is filled with a continuous flow of 1.00(1)% $CO_2$ diluted in $N_2$ at a total pressure of 750(4) Torr.

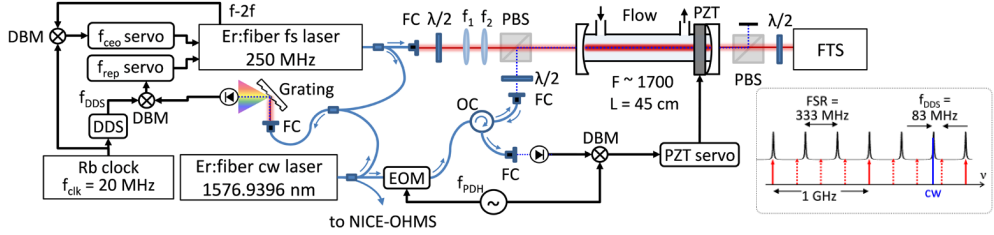

Fig. 1. Experimental setup: f-2f, beat note of the f-2f interferometer; DBM, double-balanced mixer; $f_{clk}$, Rubidium clock frequency; DDS, digital direct synthesizer; $f_{DDS}$, DDS frequency; FC, fiber collimator; $\lambda/2$, half-waveplate; $f_{1,2}$, mode-matching lenses; PBS, polarizing beam splitter; PZT, piezoelectric transducer; FTS, Fourier transform spectrometer; EOM, electro-optic modulator; OC, optical circulator; $f_{PDH}$, PDH modulation frequency. Inset: the matching of the comb (red) and the cw laser (blue) to the cavity (black). The transmitted comb lines are shown by the red solid arrows.

The comb source is an amplified Er:fiber femtosecond laser emitting in the 1505 – 1650 nm range with a repetition rate, $f_{rep}$, of 250 MHz and ~400 mW of optical power. The comb offset frequency, $f_{ceo}$, is stabilized to 20 MHz by locking the beat signal of the f-2f interferometer to a frequency generated by a GPS-referenced Rb oscillator via feedback to the current of the oscillator pump diode. The comb repetition rate is stabilized by locking a beat note between the cw laser and the nearest comb line to a frequency reference provided by a tunable direct digital synthesizer (DDS) referenced to the Rb oscillator. To measure the beat note, the cw and comb beams are combined using a fiber coupler and coupled into free space, where the light is dispersed by a grating and imaged onto a detector. The error signal is obtained by mixing the reference frequency, $f_{DDS}$, with the measured beat note, and the resulting correction signal is fed to the oscillator intracavity PZT and EOM to control $f_{rep}$. The DDS frequency is set to around 83 MHz to bring every fourth comb line close to every third cavity resonance (see inset in Fig. 1). Because of the different spacing of the comb lines and the cavity modes, the repetition rate of the comb light transmitted through the cavity is 1 GHz. The comb beam is mode-matched to the $TEM_{00}$ modes of the cavity using two lenses. The lens positions and beam alignment are optimized by scanning the cavity length, observing the transmission through the cavity using a photodiode placed directly after the cavity, and minimizing the amplitude of the higher order transverse modes. This implies that the mode-matching is optimized for the wavelengths at which the comb power is maximized. The comb and cw laser beams are combined in free space on a polarizing beam splitter (PBS) cube and coupled into the cavity with orthogonal polarizations. The beams behind the cavity are separated using a second PBS cube. Because of the finite extinction ratio of the PBS cubes a small fraction of the cw beam leaks through them. The cavity transmission is measured using a fast-scanning FTS with a nominal resolution of 1 GHz (equivalent to an optical path difference, OPD, of 30 cm). Matching the nominal resolution to the repetition rate of the transmitted comb allows measurements of the comb line intensities without the influence of the instrumental lineshape [18]. The comb interferogram is measured with an auto-balancing InGaAs detector similar to the one described in ref. [23] and the OPD is calibrated using the interferogram of a co-propagating HeNe cw laser beam, measured with an Si detector. Both interferograms are measured with a 20-bit data acquisition card with a sampling rate of 5 Msample/second.

To record the transmission spectrum of the cavity, the comb lines are scanned across the cavity modes by tuning the DDS frequency, $f_{DDS}$, in 120 steps of 20 kHz. For each step, 20 interferograms with 1 GHz nominal resolution are recorded. A single interferogram is acquired in 0.6 s (which includes 0.2 s dead time needed for the FTS mirror to turn around) resulting in a total acquisition time of ~24 minutes. The transmission spectrum for each $f_{DDS}$ value is retrieved from the average of the fast Fourier transforms (FFT) of the 20 individual interferograms. The final cavity mode spectrum is obtained by interleaving spectra measured with different $f_{DDS}$ values. For determination of the comb line frequencies, $f_{rep}$ is monitored during the measurement using a frequency counter.

As the comb lines are scanned across the cavity modes the total power on the auto-balancing detector in the FTS changes by more than two orders of magnitude. This detector shows a slight nonlinearity at low optical powers, caused by the feedback circuit. We characterized this nonlinearity by measuring the peak-to-peak value of the interferogram and the corresponding DC level of the signal photodiode as a function of optical power. We fitted a 4$^{th}$ order polynomial to the relation between the DC levels and the peak-to-peak values of the interferogram. This function was subsequently used to correct the intensity of the interferograms before taking the FFT. No other corrections were made in the interleaving process.

## 4. Results

Figure 2 shows a cavity transmission spectrum containing 15000 cavity modes spanning 15 THz of bandwidth between 1505 and 1650 nm. The amplitude of the spectrum is normalized to its maximum at 1600 nm and the spectral envelope follows that of the comb power spectrum. The positions of the three strongest absorption bands of $CO_2$ are indicated with blue bars and the insets show zoomed views of two $CO_2$ absorption lines from each band, where the black vertical lines are the individual cavity modes. The narrow peak at 1576.9 nm originates from the cw laser light that leaks into the FTS.

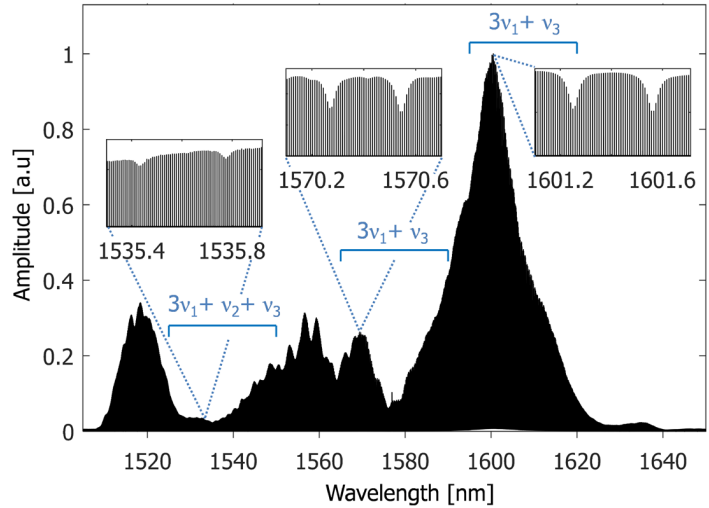

Fig. 2. A cavity transmission spectrum, where the positions of the three strongest $CO_2$ absorption bands are indicated with blue bars. Insets: parts of the spectrum containing two $CO_2$ absorption lines from each band. The amplitude of the spectrum follows that of the comb power envelope.

A Lorentzian profile was fitted to each cavity mode with the center frequency, $\nu_k$, the width, $\Gamma_k$, and the amplitude, $P_k$, as fitting parameters. The upper windows of Fig. 3 show two cavity modes within the $3\nu_1+\nu_3$ $CO_2$ band centered at ~1610 nm [for which the optical power, and thus the signal-to-noise ratio (SNR), are highest] located at (a) 1603 nm and (b) 1611 nm. Note that these modes are not resonant with $CO_2$ absorption lines. The experimental data are shown by black markers while the red curves show the fits, and the residuals are displayed in the lower windows. The average precision of the retrieved cavity mode parameters within this $CO_2$ band is of the order of 30 Hz for the center frequency, 10 Hz for the mode width, and 0.02% for the amplitude. For the cavity mode at 1611 nm the residual is almost flat, while for the cavity mode at 1603 nm, there is a small remaining structure indicating that the mode profile is not purely Lorentzian. In general, the quality of the fit varies across the cavity transmission spectrum. Figure 4(b) shows the residuals of the Lorentzian fits to cavity modes at different wavelengths (y-axis, in 2 nm intervals) as a function of the $f_{DDS}$ step number. Figure 4(a) shows the corresponding peak-to-peak voltage of the interferogram at the output of the auto-balancing detector. The color in panel (b) indicates the amplitude of the residuals (in the same units as in Fig. 2 and Fig. 3) and the black circular markers show the positions of the centers of the cavity modes with respect to the scan. The centers of the two modes shown in Fig. 3 are marked by white squares. The positions of the three $CO_2$ bands are indicated with black bars in the right margin. The residuals are flat for almost the entire $3\nu_1+\nu_3$ band centered at ~1610 nm but slightly structured for the cavity modes whose center is measured when the intensity on the detector is highest. This can be caused either by imperfect correction of the detector nonlinearity or by the fluctuations of the laser spectral envelope from step to step, which are not corrected in the interleaving process (note the vertical stripes visible e.g. at steps 55 and 65). The residuals for the $3\nu_1+\nu_3$ band centered at ~1575 nm are in general flat, however, for the cavity modes around 1577 nm the fit quality is noticeably worse because of the distortion caused by the instrumental lineshape (ILS) of the cw laser peak. This ILS is present because the sub-nominal resolution procedure is set to sample the comb lines spaced by 1 GHz, whereby the cw peak, which is separated by 333 MHz from the nearest transmitted comb line, is sampled incorrectly. Finally, the fits in most of the region around the $3\nu_1+\nu_2+\nu_3$ band centered at ~1537 nm are visibly distorted by a transverse mode appearing because of the deteriorated mode-matching at these wavelengths (the blue spot around step 60). This phenomenon is also visible at the long wavelength edge of the spectrum above 1620 nm.

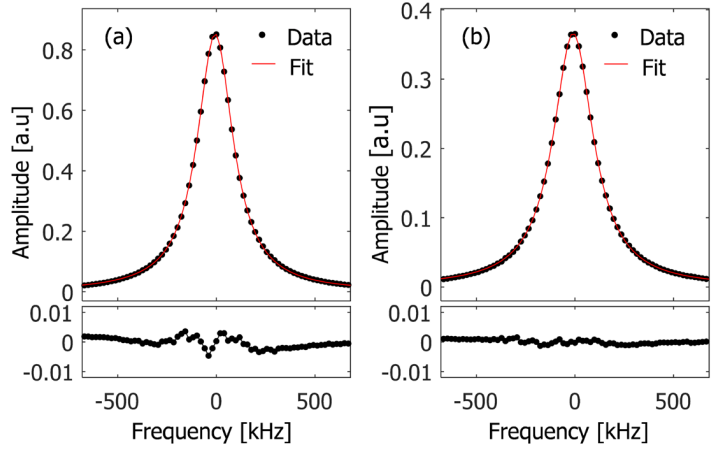

Fig. 3. Cavity modes (black markers) at (a) 1603 nm and (b) 1611 nm together with fits of a Lorentzian function (red curves), and the residuals (lower windows).

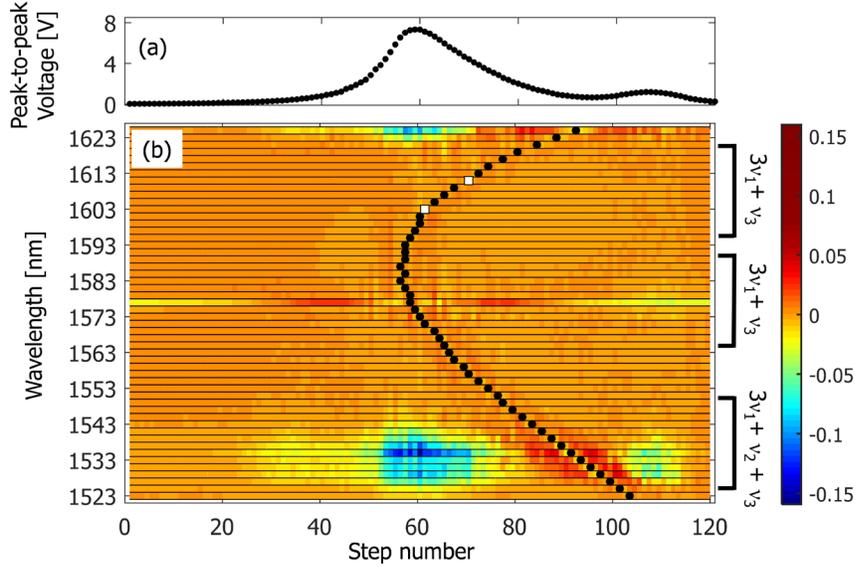

Fig. 4. (a) Peak-to-peak voltage of the interferogram at the output of the auto-balancing detector for the 120 $f_{DDS}$ steps needed to scan across the cavity mode profiles. (b) Residuals of Lorentzian fits to cavity mode profiles at different wavelengths (y-axis, in 2 nm intervals) as a function of the $f_{DDS}$ step number. The color indicates the amplitude of the individual residuals (in the same units as in Fig. 3) and the black markers indicate the position of the centers of the cavity modes with respect to the scan. The white squares mark the positions of the cavity modes shown in Fig. 3.

Figures 5(a) and (b) show the cavity mode frequencies and widths, respectively, obtained from the Lorentzian fits to all cavity modes. Both curves comprise a sum of a slowly varying baseline (for clarity, a linear frequency grid spaced by the mean value of the cavity FSR has been subtracted from the fitted center frequencies), and the resonant contributions caused by the dispersion and absorption of the $CO_2$ bands. [The dispersion of the $3\nu_1+\nu_2+\nu_3$ band at ~1537 nm, which has transition linestrengths one order of magnitude lower than the other two bands, is shown in the inset of Fig. 5(a) after another linear baseline has been removed]. Note that the baselines in the mode frequency and width spectra are significantly less structured than the envelope of the cavity transmission spectrum, Fig. 2. This is because the former are slowly varying functions of frequency, determined by the frequency dependence of the non-resonant index of refraction and the dispersion and reflectivity of the cavity mirrors, but are independent of the comb power. The baseline of the mode width spectrum is distorted below 1540 nm because of the higher-order transverse modes that affect the width retrieved from Lorentzian fits to the $TEM_{00}$ modes. The peak around 1577 nm is caused by the cw laser light leaking through the cavity.

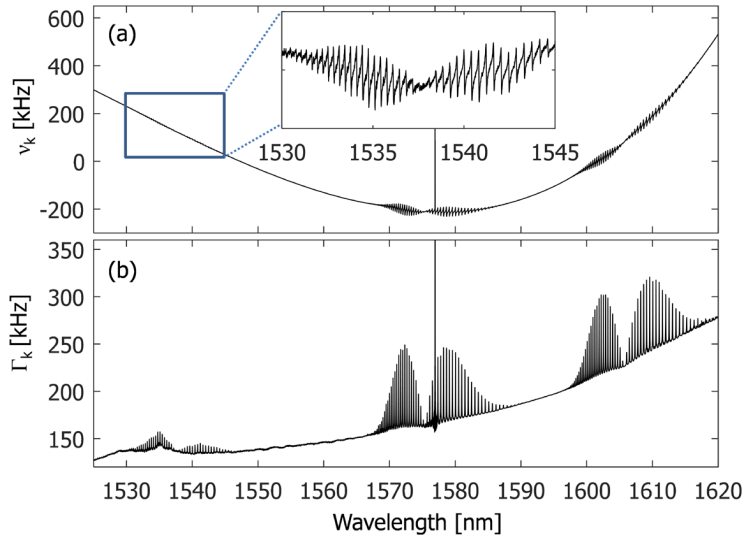

Fig. 5. (a) Cavity mode frequencies, after a linear grid spaced by the mean value of the FSR has been subtracted. Inset: zoomed view of the $3\nu_1+\nu_2+\nu_3$ $CO_2$ band centered at ~1537 nm after another linear baseline has been subtracted. (b) Cavity mode widths.

We analyze the spectra of the three absorption bands separately. Below we first present in detail the data analysis for the $3\nu_1+\nu_3$ band of $CO_2$ centered at ~1610 nm. Figure 6(a) shows the mode frequency shift, $\nu_\varphi$ (black markers, left y-axis), caused by the dispersion (right y-axis) of the $3\nu_1+\nu_3$ band of $CO_2$, retrieved from the data shown in Fig. 5(a), together with a fit of a model based on Eqs (4) and (9) (red curve). The slowly varying baseline originating from the non-resonant intracavity dispersion [see Fig. 5(a)] and etalon fringes were removed by fitting a sum of a third-order polynomial function and low frequency sine functions together with the molecular model and subtracting it from the data. Molecular dispersion was calculated using the imaginary part of the complex Voigt profile with the Doppler width ranging between 344 and 349 MHz and line parameters from the HITRAN 2016 database [24]. The non-resonant refractive index, $n$, was assumed to be equal to 1.00027, the mean value for nitrogen in this wavelength range [25], because the precision of the fit did not allow to see the influence of the dispersion of $n$. The $CO_2$ concentration was the only fitting parameter. The residual shown in the lower window indicates an overall good agreement with the model, although there is a small remaining structure, mainly at the lower wavelengths. This correlates well with the more structured residuals of the fits to the cavity modes at the lower wavelengths, shown in Fig. 3 and Fig. 4. The returned $CO_2$ concentration is 1.0048(2)%, where the uncertainty is given by the precision of the fit. This concentration agrees with the specified concentration of the gas sample within the accuracy of the pressure measurement. The SNR for the strongest line is 850 and the noise-equivalent dispersion sensitivity is $2.2 \times 10^{-8}$ $cm^{-1}$.

The mode broadening, $\Gamma_\alpha$ (left y-axis), caused by the absorption (right y-axis) of the $3\nu_1+\nu_3$ band of $CO_2$ is shown by black markers in Fig. 6(b). The red curve shows a fit of the model based on Eqs (5) and (13) calculated using the real part of the complex Voigt profile, line parameters from the HITRAN 2016 database, $n = 1.00027$, and the $CO_2$ concentration as the only free parameter. The sum of a third-order polynomial and low frequency sine functions was fitted together with the molecular model to remove the slowly varying baseline, $\Gamma_k^0$, and etalon fringes, respectively. The returned $CO_2$ concentration from the fit is 0.9995(1)%. The residual shown in the lower window indicates an overall good agreement with the model. The different structure of the residual compared to that of the dispersion is caused by the different symmetry of the signal, as well as the fact that the determination of the mode width is affected differently than the center frequency by any deviation of the measured profile from a Lorentzian function. This we believe is one of the reasons why the concentration deviates slightly (relative deviation of 0.5%) from that retrieved from the dispersion spectrum. The SNR for the strongest line is 1500 and the absorption sensitivity is $1.2 \times 10^{-8}$ $cm^{-1}$.

Finally, the cavity mode amplitudes, $P_k$, retrieved from the Lorentzian fits are used to obtain a cavity-enhanced transmission spectrum. The comb power envelope, $P_k^0$, needed for normalization of the spectrum, is found by taking the ratio between the cavity mode amplitudes and the model based on Eqs (5) and (14), smoothening it with a moving average function, and fitting the sum of a third-order polynomial and low frequency sine functions to it. The spectrum is then divided by the fitted comb envelope to yield the cavity-enhanced transmission spectrum, $A_k^\alpha$, shown by black markers in Fig. 6(c). The red curve is a fit of the model based on Eqs (5) and (14), calculated using the real part of the complex Voigt profile and the line parameters from the HITRAN 2016 database, with the $CO_2$ concentration as the only free parameter. The cavity length, $L$, is assumed to be equal to $c/2/(4nf_{rep}/3)$, where $f_{rep}$ is the repetition rate that maximizes the transmission through the cavity. The cavity reflection coefficient is calculated using Eq. (12) from the baseline of the cavity mode width spectrum, $\Gamma_k^0$, and $R_k + T_k = 1$ is assumed. The residual of the fit is shown in the lower window, again indicating good agreement with the theoretical model. The returned $CO_2$ concentration is 1.0091(3)%, where the uncertainty is given by the precision of the fit (0.0001%) and the uncertainty of $\Gamma_k^0$ (0.00024%). We note that while quantitative analysis of cavity-enhanced transmission spectra measured using other techniques requires an independent determination of cavity length and mirror reflectivity, here these parameters are retrieved from the $f_{rep}$ value and the mode width spectrum that are measured simultaneously. However, the existence of a structured baseline, which needs to be removed in the normalization process, is a source of uncertainty in the final spectrum that is difficult to quantify. Nevertheless, the retrieved concentration deviates by less than 1% from those obtained from the CE-CRIS spectra. The SNR for the strongest line in the spectrum is 1200 and the absorption sensitivity is $1.0 \times 10^{-8}$ cm$^{-1}$.

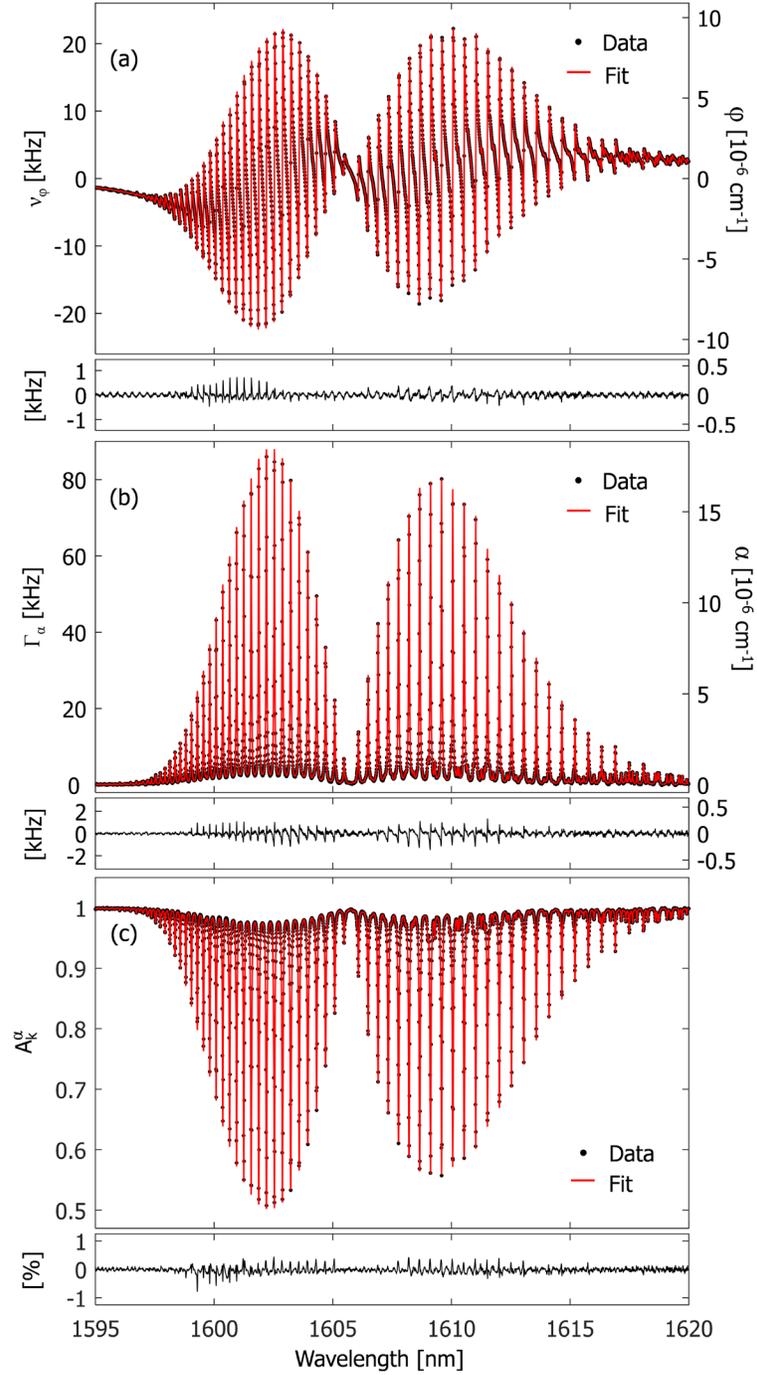

Fig. 6. Spectra of the $3\nu_1+\nu_3$ band of 1% $CO_2$ in $N_2$ at 750 Torr (black markers) with fits of the corresponding models (red curves) and residuals (lower windows). (a) Dispersion (right y-axis) retrieved from the mode frequency shift (left y-axis). (b) Absorption (right y-axis) retrieved from the mode broadening (left y-axis). (c) Cavity-enhanced transmission spectrum retrieved from the mode amplitude.

A similar analysis was performed for the $3\nu_1+\nu_3$ $CO_2$ band centered at ~1575 nm and the $3\nu_1+\nu_2+\nu_3$ band centered at ~1537 nm, and the results are presented in Fig. 7 and Fig. 8, respectively. The panels (a) and (b) show the mode frequency shift and broadening (left y-axes), revealing the dispersion and absorption spectra (right y-axes), respectively, while the panels (c) show the cavity-enhanced transmission spectra. The data are shown by black markers, fits of the corresponding models by the red curves, and residuals are shown in the lower windows.

In Fig. 7, the distortion caused by the ILS of the cw laser peak at ~1577 nm is visible in all spectra of the $3\nu_1+\nu_3$ band. To remove its influence on the fit, a small range around this wavelength is not considered during the fitting. The SNR is above 700 for all three spectra and the returned $CO_2$ concentrations from the fits are 0.9821(1)%, 0.9849(2)%, and 0.9994(6)% for dispersion, absorption and cavity-enhanced spectra, respectively. While the concentrations retrieved from fits to dispersion and absorption spectra agree with each other to within 0.2%, they deviate from the specified $CO_2$ concentration and the concentration retrieved from fits to the $3\nu_1+\nu_3$ band at ~1610 nm by ~2%. This discrepancy can be caused by inaccuracy of linestrengths reported in the HITRAN database for this absorption band – a similar discrepancy was observed for the R16e line in this band in ref. [18]. This issue will be a subject of future investigation. The concentration retrieved from the cavity-enhanced transmission spectrum is higher than those from the other two spectra, which we believe might be caused by an error introduced during the baseline removal process. It should be noted though that all spectra measured at this wavelength range are affected to a different extent by the ILS of the cw laser peak. To remove its influence, polarization components with better extinction ratios need to be used or another cw laser wavelength should be chosen for the cavity length stabilization.

In Fig. 8(a), the dispersion spectrum of the $3\nu_1+\nu_2+\nu_3$ band is offset from zero by the wing of the dispersion of the $3\nu_1+\nu_3$ band at ~1575 nm, which is taken into account in the model. The concentration returned from the fit is 0.999(1)%, in agreement with that obtained from the $3\nu_1+\nu_3$ band at ~1610 nm. The cavity mode width spectrum, shown in Fig. 8(b), however, is clearly distorted by the presence of the higher-order transverse cavity modes (see Fig. 4 and Fig. 5). The line intensities in the R- and P-branches are higher and lower, respectively, than the model fitted to both branches simultaneously. The concentration retrieved from the fit to both branches is 1.247(3)%, while a fit to only the P-branch returns 1.069(5)%, closer to the expected value. This correlates well with the fact that the R-branch is more affected by the presence of the transverse mode. To obtain the cavity mirror reflectivity needed to analyze the cavity-enhanced transmission spectrum, shown in Fig. 8(c), we extrapolated the baseline of the cavity mode width spectrum across the R-branch using a linear fit to the baseline of the P-branch. The concentration returned from the fit to this spectrum is 1.020(3)%. The accuracy of the measurement of this $CO_2$ band will be improved when the mode-matching is optimized (or the transverse mode is taken into account in the fitted model) and the comb spectral envelope is optimized to increase the power in this wavelength range (the SNR is of the order of 100 for this band because of the lower comb power and weaker molecular linestrengths).

## 5. Conclusions and outlook

We have demonstrated for the first time broadband cavity-enhanced complex refractive index spectroscopy (CE-CRIS) and retrieved the dispersion and absorption spectra of three bands of $CO_2$ in the range between 1525 nm and 1620 nm from a direct measurement and characterization of the transmission modes of a Fabry-Perot cavity filled with the gas sample. The cavity transmission spectrum spanning 15 THz (limited only by the spectral bandwidth of the comb) was measured using a comb-based Fourier transform spectrometer with sub-nominal resolution, which enables precise measurement of thousands of cavity modes simultaneously with absolute frequency calibration provided by the comb itself. The cavity mode parameters were obtained from fits of Lorentzian profiles. The cavity mode shift and broadening are directly proportional to the molecular dispersion and absorption, respectively, and independent of the cavity parameters, such as the length and mirror reflectivity. This makes the technique calibration-free and less sensitive to systematic errors. Hence, broadband CE-CRIS combines the advantages of cw-laser-based 1D-CMDS [13-15] and CMWS [12] with orders of magnitude wider spectral bandwidth, which allows measurements of the complex refractive index of several molecular bands simultaneously. Moreover, the third cavity mode parameter, i.e. the amplitude, gives, after normalization with the comb envelope, the cavity-enhanced transmission spectrum. Such transmission spectra measured using other techniques require calibration of cavity finesse for retrieval of quantitative information; here, however, the cavity mirror reflection is determined from the baseline of the cavity mode width spectrum, which eliminates the need for a separate finesse measurement and makes this spectrum calibration-free as well.

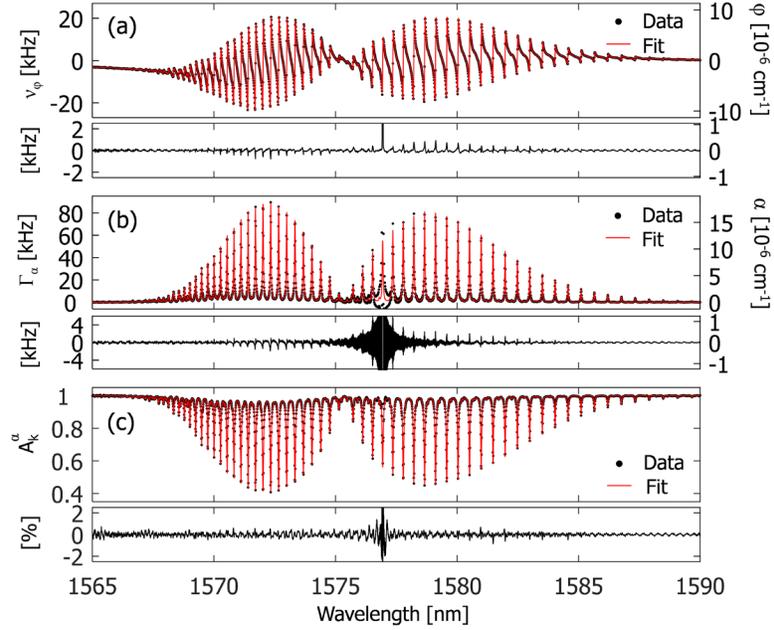

Fig. 7. Spectra of the $3\nu_1+\nu_3$ absorption band of 1% $CO_2$ in $N_2$ at 750 Torr (black markers) retrieved from the (a) mode frequency shift, (b) cavity mode broadening, and (c) mode intensities, together with fits (red curves) and residuals in the lower windows.

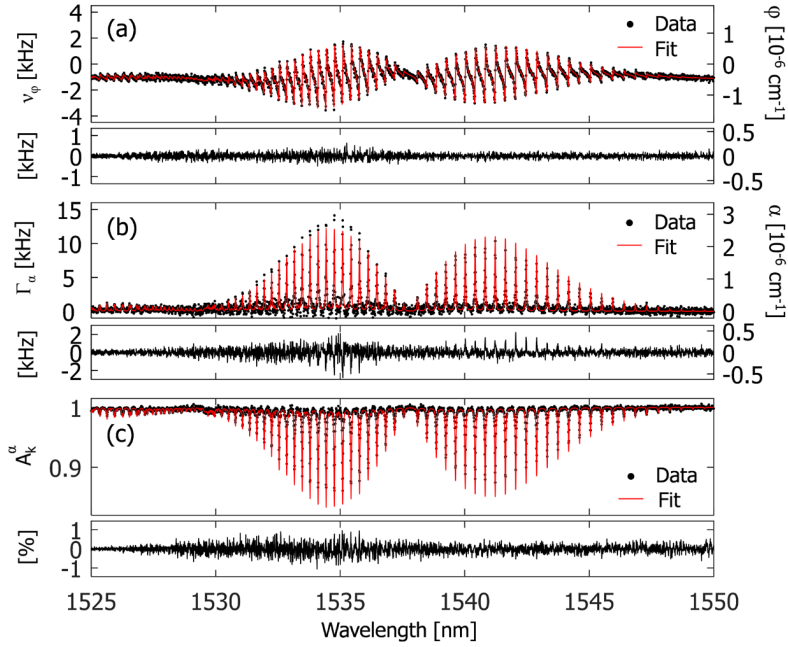

Fig. 8. Spectra of the $3\nu_1+\nu_2+\nu_3$ absorption band of 1% $CO_2$ in $N_2$ at 750 Torr (black markers) retrieved from (a) the mode frequency shift, (b) cavity mode broadening, and (c) mode intensities, together with fit (red curves) and residuals in the lower windows.

The quality of the Lorentzian fits to the cavity modes varies across the spectrum because of the different optical power of the comb lines, deterioration of the mode-matching at the edges of the spectrum, and leakage of the cw laser light into the FTS; thus the overall precision of the fits will be improved when these issues are addressed. The signal-to-noise ratio in the CE-CRIS spectra is up to 1500. This is lower than in the best demonstrations of CE-CRIS using cw lasers [14], but is expected to improve when the cavity finesse is increased. Moreover, the sampling point spacing in the CE-CRIS spectra obtained from a single cavity transmission measurement is given by the repetition rate of the transmitted comb. Denser sampling point spacing can be obtained by performing measurements with different cavity lengths (which shifts the positions of the cavity modes) and interleaving the spectra. This will allow measurements of complex refractive index at lower sample pressures than demonstrated here.

Compared to other comb-based cavity-enhanced techniques, broadband CE-CRIS is the only one with spectral bandwidth and frequency calibration given solely by the properties of the comb. In cavity-enhanced optical frequency comb Fourier transform spectroscopy [23, 26], in which the laser is tightly locked to the cavity, cavity mirror dispersion limits the transmitted bandwidth and introduces an asymmetry in the absorption lines, which has to be taken into account in the quantitative analysis of the spectra [26]. In VIPA-based systems, the simultaneous transmitted bandwidth is limited by the detector array size [27]. In continuous-filtering Vernier spectroscopy [28], the bandwidth is determined by the comb, but the technique requires calibration of the frequency scale and spectra are convolved with an instrumental line shape.

In conclusion, broadband CE-CRIS using a comb-based Fourier transform spectrometer with sub-nominal resolution is an excellent tool for calibration-free simultaneous measurements of the real and the imaginary parts of the molecular refractive index with spectral bandwidth and frequency scale given by the comb. Measurements of complex refractive index as a function of pressure will allow multispectral fitting for line parameter retrieval [29], opening up for precision spectroscopy of several bands simultaneously. This will allow e.g. verification of vibrational and rotational molecular energy levels with improved accuracy and investigation of line-mixing effects. Since the absorption and dispersion spectra are measured with high precision under the same experimental conditions they can provide new information about the relation between the real and imaginary parts of the molecular line shapes [15] and allow assessment of the far wing contributions of atmospheric species for radiative transfer calculations [30].

**Funding**